
\magnification=\magstep1
\hsize=15.6 truecm
\vsize=21.5 truecm
\baselineskip=14 pt
\null \vfill
{\sevenrm\hbox to 15truecm{ISSN 0133--462X\hfil ITP
 Budapest Report No. 502}
\hbox to 15truecm{\hfil July 1993}}

\centerline{$\phantom{\pi}$}

\vskip 1.5truecm

\centerline{\bf A new family of $SU(2)$ symmetric integrable
sigma models}

\vskip 4 truecm

\centerline{J. Balog, P. Forg\'acs }
\centerline{Research Institute for Particle and Nuclear Physics}
\centerline{H-1525 Budapest 114, P. O. B. 49, Hungary}
\medskip
\centerline{Z. Horv\'ath, L. Palla}
\centerline{Institute for Theoretical Physics}
\centerline{Roland E\"otv\"os University}
\centerline{H-1088 Budapest, Puskin u. 5-7, Hungary}
\vskip 4.5truecm

\centerline{{\bf Abstract}}

Local Lagrangeans are derived for a class of $SU(2)$ invariant
sigma models admitting two commuting Kac-Moody algebras at the
level of Poisson brackets. The one loop renormalizability of
these models is established. Some heuristic arguments are
presented in favour of their quantum integrability.
\vfill
\eject

\noindent
In this paper we study a special class of two dimensional
sigma models which admit two Poisson bracket (PB) commuting
Kac-Moody algebras (KMA's). To consider such models has
been suggested some time ago by Rajeev [1] whose motivation
was to establish a direct connection between the principal
chiral sigma model (P$\Sigma$M) and the chiral Gross-Neveu
(or non-abelian Thirring) model. This approach might be
useful to provide an alternative quantization of the
P$\Sigma$M, see refs.\ [2,3] for the `standard' methods.

One of our main result is a family of {\it local}
Lagrangeans depending on a `deformation' parameter
(in addition to the usual coupling) providing a realization
of the canonical PB's corresponding to the commuting KMA's.
Our Lagrangean describes a deformation of the P$\Sigma$M
in the sense that for some value of the deformation parameter
it yields the standard eqs.\ of motion and PB's for the
currents, however, it cannot be transformed to the known one
by local field redefinitions.\hfill\break
We also show that the deformed Lagrangeans are renormalizable
at the one loop level and provide some heuristic arguments about
their quantum integrability.

Let us start with the standard equations
of motion written in terms
of currents valued in the Lie algebra of a semisimple
Lie group $G$, $I_{\mu }={\rm
i}\lambda ^aI_{\mu }^a$: \footnote{*}{Our conventions are:
$\mu ,\nu =0,1$; $\gamma ^{00}=-\gamma ^{11}=1$; $v_{\pm
}=v_0\pm v_1$; $\epsilon ^{01}=1$; $[\lambda ^a,\lambda ^b]={\rm
i}f^{abc}\lambda ^c$ with $f^{abc}$ being the structure
constants of $G$}
$$\eqalignno{
&\partial _{\mu }I^{\mu }=0 &(1a) \cr
&\partial _\mu I_\nu -\partial
_\nu I_\mu +[I_\mu,I_\nu]=0 &(1b)
\cr}$$
As it is well known eqs.\ (1a,1b) are the compatibility
conditions of an auxiliary linear problem [4]:
$$
[{\cal D}_{\mu}(\lambda),{\cal D}_{\nu}(\lambda)]=0
\qquad
{\cal D}_{\mu}(\lambda)=\partial_\mu +
{\lambda^2\over\lambda^2
-1}I_\mu + {\lambda\over\lambda^2-1}\epsilon_{\mu\rho}I^{\rho}
\eqno(1c)
$$
hence they admit an infinite number of (classically) conserved
quantities. Most known two dimensional integrable models
correspond to eqs.\ (1) or to some of their reductions.

The direct application of the quantum inverse scattering methods
[5] to eqs.\ (1) has been severely hampered by the
non-ultralocality of many interesting models (e.g.\ the P$\Sigma$M
and most of its reductions).
A promising alternative to the quantization of models in this
class seems to be the use of the conserved non-local charges
derived from (1). This approach has been pioneered by
L\"uscher [6] and developed further in refs.\ [7,8].
In ref.\ [8] a general framework (called massive current
algebras) has been set up, requiring in particular that the
currents $I_\mu$ satisfying the quantum version of eqs.\
(1a, 1b) be Kac-Moody currents in the ulraviolet limit.
Models with massive current algebras are quantum integrable
i.e.\ they possess an infinite number of conserved
(non-local) charges implying the absence of particle
production and the factorizability of the $S$-matrix.
To show that a given model has a massive current algebra
is a non trivial problem requiring non-perturbative methods.
In particular the P$\Sigma$M does not fit in this framework.
We hope that our deformed sigma models with the commuting KMA's
are in this class.

The principal sigma model in its standard form is obtained by
solving
eq.\ (1b) with $I_\nu=g^{-1}\partial _\nu g$
($g\in G$) and then eq.\ (1a) is obtained
from the Lagrangean
$$S_1=\int d^2\xi
{\cal L}_1={1\over 2e^2}\int d^2\xi {\rm Tr}\partial _\mu g^{-1}
\partial ^\mu g \eqno(2)$$
(Here we introduced a coupling constant, $e^2$, that could be
scaled to one classically, but we prefer to keep it as it will
play an important role in the quantum theory.)
However in a completely analogous way we can solve eq.\ (1a) with
$I^\mu =\epsilon ^{\mu \nu }\partial _\nu \Phi$,
then eq.\ (1b) for $\Phi $ is nothing
but the equation following from the variation of
$$S_2=\int d^2\xi
{\cal L}_2={1\over 2e^2}\int d^2\xi \bigl( \partial _\mu \Phi
^a\partial ^\mu \Phi ^a-{1\over 3}f^{abc}\epsilon ^{\mu \nu}
\Phi ^a\partial _\mu \Phi ^b\partial _\nu \Phi ^c\bigr) \eqno(3)$$
(This model was briefly discussed in [9].) The models described
by $S_1$ and $S_2$ are physically different, since
not only the PB structure of the currents is completely different
but also -- as we
shall see later -- $S_2$ describes a non asymptotically free
theory, while the P$\Sigma$M is asymtotically free.

In both models the Hamiltonian and the momentum can be written in terms of
the currents as
$$H=\int d\xi ^1{\cal H}\qquad {\cal H}={1\over 2}(V+\bar V)\qquad
V={1\over 2e^2}I^a_+I^a_+\eqno(4a)$$
$$P=\int d\xi ^1{\cal P}\qquad {\cal P}={1\over 2}(V-\bar V)\qquad
\bar V={1\over 2e^2}I^a_-I^a_-\eqno(4b)$$
but of course the Poisson algebra among the currents, $I^a_\mu$,
following from the canonical PB's defined by
${\cal L}_1$ and ${\cal L}_2$ is different in the two cases.

In ref.\ [1] it has been pointed out, that in the case of the
P$\Sigma$M the Poisson algebra of the currents admits an
appropriate
deformation depending on a continuous parameter such that the
Hamiltonian
equations of motion for the standard Hamiltonian, eq.\ (4a), are
still identical to eqs.\ (1a, 1b). We generalise this observation
slightly by allowing the presence of a Wess-Zumino
term with a general coefficient ($\rho \ne 0$)
in ${\cal L}_1$, i.e.\ eqs.\ (1a, 1b) modified as
$$\partial _-I_+=-{1-\rho \over 2}[I_+,I_-]\qquad \qquad
\partial _+I_-={1+\rho \over 2}[I_+,I_-]\eqno(5)$$
and claim that eqs.\ (5) are the Hamiltonian equations for $H$ in
eq.\ (4a) if the equal
time PB's among the currents have the following general form:
$$\{ I^a_+(\sigma ),I^b_+(\hat \sigma )\}=e^2f^{abc}\bigl( (1-\rho)
I^c_-(\sigma )-[(1+2x)-\rho (2x-1)]I^c_+\bigr) \delta
(\sigma-\hat \sigma)+2e^2\delta ^{ab}\delta ^{'}(\sigma -\hat
\sigma )$$
$$\{ I^a_-(\sigma ),I^b_-(\hat \sigma )\}=e^2f^{abc}\bigl( (1+\rho)
I^c_+(\sigma )-[(1+2x)+\rho (2x-1)]I^c_-\bigr) \delta
(\sigma-\hat \sigma)-2e^2\delta ^{ab}\delta ^{'}(\sigma -\hat
\sigma )$$
$$\{ I^a_+(\sigma ),I^b_-(\hat \sigma )\}=-e^2f^{abc}\bigl( (1-\rho)
I^c_+(\sigma )+(1+\rho )I^c_-\bigr) \delta
(\sigma-\hat \sigma)\eqno(6)$$
Here $x$ is a new constant parameter that does not appear in
eqs.\ (5), but whose presence in eq.\ (6) is allowed by the Jacobi
identity. (Setting $\rho =0$ in eq.\ (6) we recover the
Poisson algebra suggested in ref.\ [1].)
Since in the Hamiltonian framework the specification
of the fundamental PB's forms an essential part of the
definition of the model, giving eq.\ (4-6) we defined a whole
family of models depending on the new parameter $x$.
Setting $x=1$ (and $\rho =0$) in eq.\ (6) we get the usual PB's
of the principal sigma model while the
$x=-1$ (and $\rho =0$) case of eq.\ (6) are identical to
the canonical PB's of
$I^a_\mu$ following from ${\cal L}_2$.

The very appealing property of the $x\ne \pm 1$ ($\rho
=0$) models -- that motivated the study of ref.\ [1] -- is
that at the PB level they admit two
commuting Kac Moody algebras. Indeed if $x^2>1$ then --
as a straightforward calculation shows -- the linear
combinations
$$u^a_\pm (\sigma )={1\over 2e^2}\bigl[ -{1\over x+1}I^a_0
(\sigma )\pm
{1\over \sqrt{x^2-1}}I^a_1(\sigma )\bigr]\eqno(7)$$
form two commuting KM algebras with centres $k=\pm {1\over 2e^2}
{1\over x+1}{1\over \sqrt{x^2-1}}$ respectively, as a consequence
of eq.\ (6). (If $x^2<1$ then we have to form complex linear
combinations of $I^a_0$ and $I^a_1$ to get the KM currents
$u^a_\pm $;
$k$ in this case becomes $\pm {1\over 2e^2}
{1\over x+1}{1\over \sqrt{1-x^2}}$.) We see from these formulas
that from the point of view of the KMA the $x=\pm 1$
models are certain singular limits of the generic case.

The study of these interesting models has been severely hindered
by the fact that the Lagrangean found
in ref.\ [1] was non local and not
manifestly Lorentz invariant.
Since for the $x=\pm 1$, $\rho =0$ cases
local and manifestly Lorentz invariant
Lagrangeans exist it seems natural to ask if
this remains so for the
general $x\ne \pm 1$ case, which in a certain sense, would
interpolate between the P$\Sigma $M and the theory defined by
${\cal L}_2$ (for $\rho =0$ at least). We are going to
show that this is indeed the case at least for $G=SU(2)$.

As both ${\cal L}_1$
and ${\cal L}_2$ belong to the general class of Lagrangeans for
bosonic sigma models with torsion we are going to work in
this framework, i.e.\ we shall look for Lagrangeans of
the form
$${\cal L}={1\over2e^2}G_{AB}(X)\partial ^\mu X^A\partial _\mu X^B
-{1\over2e^2}B_{AB}(X)\epsilon ^{\mu \nu }\partial _\mu X^A
\partial _\nu X^B\eqno(8)$$
where $G_{AB}(X)$ is a metric on the underlying
manifold and the antisymmetric
tensor field $B_{AB}(X)$ is the torsion potential: $2T_{ABC}=
\partial _AB_{BC}+{\rm cyclic}$.

Together with the generalized sigma-model form of the
Lagrangean (8)
we make the following ansatz of the currents:
$$
I^a_+=Q^a_A\,\partial_+ X^A\qquad\qquad
I^a_-=R^a_A\,\partial_- X^A\eqno(9)$$
and require that
the Hamiltonian and momentum take the quadratic form (4) and that
the KM-type current algebra relations (6) are consequences
of the usual canonical Poisson brackets among the canonical
variables $X^A$ and $\partial_0\,X^A$.

Note that both $Q^a_A$ and $R^a_A$ are vielbeins corresponding to
the target space metric $G_{AB}$. It will turn out to be
convenient to introduce the corresponding matrix valued 1-forms:
$$
-iQ^a_A\, dX^A\,\lambda^a=Q\qquad\qquad\qquad
-iR^a_A\, dX^A\,\lambda^a=R\eqno(10)$$
We will make the additional, but natural assumption that
these 1-forms are related by a similarity transformation
$$
R=\epsilon\,gQg^{-1}\eqno(11)$$
where $\epsilon^2=1$ and $g$ is a group-valued matrix.

Using the eqs.\ (1a, 1b) and the canonical structure that follows
from the Lagrangean (8) with the definitions (9) we find that the
requirements (4) and (6) are equivalent to a set of
algebraic and differential equations
satisfied by the Lie-algebra valued 1-form $Q$ and group-valued
scalar matrix $g$:
$$\eqalign{
dQ&=AQ^2+\kappa\epsilon g^{-1}Q^2g\cr
dR&=BR^2+\lambda\epsilon gR^2g^{-1}\cr
\noalign{\smallskip}
T&=A{\rm Tr}\{Q^3\}+3\kappa\epsilon{\rm Tr}\{gQg^{-1}Q^2\}\cr
&=-B\epsilon{\rm Tr}\{Q^3\}-3\lambda{\rm Tr}\{g^{-1}QgQ^2\}\cr}
\eqno(12)$$
where $Q$ and $R$ are related by (11) and the torsion
3-form $T=T_{ABC}dX^AdX^BdX^C$ in (12) satisfies
$$
dT=0.\eqno(13)$$ The constants $A,B,\kappa,\lambda$ are
related to $x$ and $\rho$ in (6)
as follows:
$$\eqalign{
\kappa&={1-\rho\over2}\qquad\qquad\qquad
A=-\lambda-2x\kappa\cr
\lambda&={1+\rho\over2}\qquad\qquad\qquad
B=-\kappa-2x\lambda\cr}$$
and $\epsilon^2=1$.

The Lagrangeans ${\cal L}_1$ and ${\cal L}_2$ correspond to
special solutions of (12-13). We do not know whether there are
new, `interpolating' solutions in the general case, but we have
studied the simplest case of $SU(2)$ symmetry in detail.

We have taken the following special $SU(2)$ ansatz:
$$\eqalign{
g&={\rm cos} \phi+{\rm sin}\phi\cdot n\cr
Q&=d\psi\cdot n+\alpha\cdot dn+\beta\cdot ndn\cr}\eqno(14)$$
where $n=in^a\tau^a$; $n^an^a=1$ and $\psi$, $\phi$, $\alpha$ and
$\beta$ are functions of the single `radial' variable $r$.
(This ansatz corresponds to an $SU(2)$ symmetric metric
on the target space, the symmetry acting on $S^2$ spheres
in the usual way.)

For this 3-dimensional target space (13) is satisfied
automatically and (12) reduce to a set of ordinary
differential equations for the unknown functions
$\psi$, $\phi$, $\alpha$ and $\beta$, which can be
solved completely. In addition to the cases ${\cal L}_1$ and
${\cal L}_2$, which are also included in (14), we find various
other solutions. There are solutions with $x=1$ and $\rho
={\rm arbitrary}$ or $x={\rm arbitrary}$ and $\rho=0$.
(Within our $SU(2)$ ansatz there are no solutions with both
parameters $x$ and $\rho$ taking arbitrary values, although this
would be allowed by the algebra (6).) The former case corresponds
to ${\cal L}_1$ with an additional WZ term. In our variables
this Lagrangean is (with $r=w$)
$$
{\cal L}={1\over2e^2}\Big\{\partial_\mu w\,\partial^\mu w
+{\rm cos}^2w\,\partial_\mu n^a\partial^\mu n^a-
\rho(w+{\rm sin}w\,{\rm cos}w)\epsilon^{abc}\epsilon^{\mu\nu}
n^a\partial_\mu n^b\partial_\nu n^c\Big\}
\eqno(15)$$
The new solutions correspond to
$\rho=0$ and after some linear rescaling they can be
represented as
$$
{\cal L}={1\over 2e^2}\Big\{ \partial_\mu r\partial^\mu r+
{\beta_0\over x+1}\partial_\mu n^a\partial^\mu n^a+
{r-\alpha_0\over x+1}\epsilon^{abc}\epsilon_{\mu\nu}
n^a\partial_\mu n^b \partial_\nu n^c\Big\}
\eqno(16)$$
and
$$\eqalign{
I^a_+=E^a_+\qquad\qquad E&=dr\cdot n+\alpha_0\cdot dn+\beta_0
\cdot ndn\cr
I^a_-=F^a_-\qquad\qquad F&=-dr\cdot n-\alpha_0\cdot
dn+\beta_0\cdot n
dn\cr}\eqno(17)$$
where

\noindent case a)
$$
\phi={\rm const.}\not=0,\pi\qquad\qquad x=-{\rm cos}2\phi$$
$$
\alpha_0=-{1\over2}{\rm ctg}\phi\qquad
\beta_0={1\over2}
$$

\noindent case b)
$$
x^2>1\qquad\qquad r={1\over\sqrt{x^2-1}}\big({\pi\over2}-w\big)$$
$$
\alpha_0={\sqrt{x^2-1}\over x+1}{{\rm sin}w\,{\rm cos}w\over
x+{\rm cos}2w}\qquad
\beta_0={{\rm cos}^2w\over x+{\rm cos}2w}
$$

\noindent case c)
$$
x=1\qquad
\alpha_0={r\over1+4r^2}\qquad
\beta_0={2r^2\over1+4r^2}
$$

\noindent case d)
$$x^2<1\qquad\qquad  \eta^2=1\qquad\qquad r={w\over\sqrt{1-x^2}}$$
$$
\alpha_0=\sqrt{{1-x\over1+x}}{{\rm sh}w\,{\rm ch}w\over
{\rm ch}2w+\eta x}\qquad
\beta_0={1\over2}{{\rm ch}2w+\eta\over
{\rm ch}2w+\eta x}$$

We note that in case b) it is natural to assume that $w$
is an angular variable. The usual argument about the
single-valuedness of the exponentialized quantum action gives
then the same quantization of the parameter $k$ (defined
under (7)) as follows from the fact that it is the centre
of the KM algebra.

It is interesting to note that we obtain the original P$\Sigma$M
in a number of different ways. In addition to
${\cal L}_1$ the same equations of motion and canonical PB's
can be obtained
from (16) in case a) (with the special choice of $\phi=\pi/2$)
or as case c). These are inequivalent Lagrangean descriptions
of the P$\Sigma$M since there is no local transformation of
the field variables that would transform these cases of (16) to
${\cal L}_1$. We also note that the conformal WZNW model,
which corresponds to (15) with $\rho=1$, can also be obtained
from (16), case b) in the limit $x\rightarrow \infty$,
$e^2x^2\rightarrow {\rm const.}$

The next important question concerns the quantum integrability of
our models. A promising possibility would be to show the existence
of massive current algebras of ref.\ [8] i.e.\ the existence
of conserved non-local charges. This problem requires further
study. Here we merely want to
show, that an appropriate modification of Polyakov's heuristic
argument for the existence of a {\it local} higher spin
conserved quantity in the quantized versions of $O(N)$ sigma
models [10] applies to the entire family of our $SU(2)$
models as well. Since
this quantity is the quantum descendant of a local
{\it non polynomial} classical quantity [4] we derive this latter
first.

Classically, as a consequence of eqs.\ (1a, 1b), we have
$$\partial _-{\rm Tr}I_+^2=0\eqno(18)$$
This equation (together with its chiral partner) expresses the
vanishing of the trace of the energy momentum tensor thus the
classical conformal invariance of our models. It is not difficult
to establish that in the case of $SU(2)$, exploiting the specific
properties of $\epsilon ^{abc}$ together with eqs.\ (1), but
without
ever using eq.\ (6), we get
$$\partial _-{\rm Tr}(\partial _+I_+)^2=-
\partial _+\{ {1\over 2}{\rm Tr}I_+^2{\rm Tr}I_+I_-\} +{3\over 2}
{\rm Tr}I_+^2\partial _+({\rm Tr}I_+I_-)\eqno(19)$$
Introducing
$$U={\rm Tr}\Bigl( \partial _+\bigl[ {I_+\over \sqrt{{\rm
Tr}I_+^2}}
\bigr]\Bigr)^2,\qquad V={{\rm Tr}I_+I_-\over \sqrt{{\rm
Tr}I_+^2}}$$
we can indeed write this equation in the form of a non polynomial
conservation equation $\partial _-U=\partial _+V$. However the
essential part of Polyakov's argument is the assumption about the
quantum modifications of eqs.\ (18), (19) generated by the
conformal
anomaly. According to this assumption
(which is implicitly based on the existence of a local
Lagrangean) the anomalous terms appearing
in eqs.\ (18), (19) are local operators with conformal dimensions,
spins
and global quantum numbers determined by the classical equations.
Supplementing these ideas
with the additional assumption that all these
operators can be constructed from $I_+$, $I_-$ we find the quantum
version of eq.\ (18) as
$$\partial _-{\rm Tr}I_+^2=\mu (e^2,x)\partial _+{\rm Tr}I_+I_-
\eqno(20)$$
Furthermore, using the equations of motion, we can show that the sum
of all possible anomalous terms appearing in eq.\ (19) can be
written as the sum of total $\partial _+$ derivatives plus a term
proportional to ${\rm Tr}I_+^2\partial _+({\rm Tr}I_+I_-)$.
Thus integrating the
quantum version of eq.\ (19) over $x^+$ and dropping
the surface terms we get
$$\partial _-\int dx^+{\rm Tr}(\partial _+I_+)^2=(3/2+\nu )
\int dx^+{\rm Tr}I_+^2\partial _+({\rm Tr}I_+I_-)\eqno(21)$$
(In eqs.\ (20), (21) $\mu $ and $\nu $ are some numerical factors
related
to the conformal anomaly.)
Substituting $\partial _+({\rm Tr}I_+I_-)$ from eq.\ (20)
 we indeed find
$$\partial _-Q=0,\qquad Q=\int dx^+({\rm Tr}(\partial _+I_+)^2-
(1/2\mu )(3/2+\nu )({\rm Tr}I_+^2)^2)\eqno(22)$$
The exsistence of this spin three conserved quantity, $Q$,
would, according to the standard argument [10, 11] guarantee the
absence of particle production.

The advantage of using Lagrangeans that are special cases of the
general framework of sigma-models with torsion is that
we can use the renormalizability of these models. Actually, it is
not completely trivial that our models are renormalizable since
although the general models are [12], we have to show that the
neccessary counter terms, which are known in the general case
[13],
can be obtained as renormalizations of our coupling parameters
$e^2$ and $x$ (together with a non-linear renormalization of
the `radial' field $r$). At least to one-loop order, this is
indeed the case, and we obtain, using the general results
in the dimensional regularization scheme [13], that under
the renormalization group transformation of the substraction point
$\mu\rightarrow e^t\mu$ the renormalized couplings
$\alpha=e^2/\pi$ and $x$ satisfy the following RG equations
$$\eqalign{
\dot \alpha&=(1-2x)\alpha^2+{\cal O}(\alpha^3)\cr
\dot x&=(x^2-1)\alpha+{\cal O}(\alpha^2)\cr}
\eqno(23)$$
(23) is valid for all cases of the new Lagrangeans (16),
and it also applies to ${\cal L}_1$ and ${\cal L}_2$
(which correspond to the cases $x=1$ and $x=-1$, respectively).

{}From (23) we see that $x^2=1$ is a special fixed line in the
space of the two couplings. For $x=1$ we recover the well-known
asymptotically free behaviour of the P$\Sigma$M coupling,
whereas for $x=-1$ we find that the model described by ${\cal
L}_2$ is not asymptotically free. (This was first pointed out in
ref.\ [14].)
 Thus these two models describe very different physics, as
 mentioned earlier.

We note that the combination $1/2\pi k=\alpha(x+1)\sqrt{x^2-1}$,
which must be quantized, is an invariant of the RG trajectories
defined by (23). However, our perturbative results, which are
obtained for fixed $x$ and small $\alpha$, are only valid for
large $k$.

The most interesting trajectories correspond to the $x>1$ case.
In this case the trajectories lead to the UV fixed point $e^2=0$,
$x=\infty$. More precisely, $x\rightarrow\infty$, while
$e^2x^2\rightarrow{\rm const.}$ (since $k$ is fixed). Thus our
models tend to the conformal WZNW model at their UV fixed points.

We have also calculated the behaviour of the current-current
2-point function in perturbation theory for the properly
normalized Noether current
$$
N_\mu^a={-1\over e^2(x+1)}I^a_\mu\eqno(24)$$
In Euclidean space, after renormalization, we find
$$
\langle N^a_\mu(x)N^b_\nu(y)\rangle={\delta^{ab}\over\pi}
\int {d^2p\over2\pi}e^{-ip(x-y)}\Big({p_\mu p_\nu\over
p^2}-\delta_{\mu\nu}\Big)I(p)\eqno(25)$$
where
$$
I(p)={1\over\alpha(x+1)^2}+{1\over2(x+1)^2}\Big(
{\rm ln}{p^2\over\mu^2}+{\rm const.}\Big)+{\cal
O}(\alpha)\eqno(26) $$

Using (26) and the renormalization group we find that for the
trajectories ending at the WZNW fixed point
$$
I(p)\rightarrow{\rm const.}\eqno(27)$$
so from this point of view our models show fermionic behaviour.
At this point we would like to remark that the result in (26)
may be considered as a first step to calculate the operator
product expansion of two currents. If in the RG eqs.\ (23) the
higher order corrections can be shown to go zero
faster than the leading terms
then one can establish the existence
of a massive current algebra of ref.\ [8] by perturbation theory.
This would be similar to the calculation of L\"uscher [6]
who used asymptotic freedom to find the short distance expansion
of the product of two currents.

In conclusion in this letter we derived local Lagrangeans for a
new family of $SU(2)$ invariant sigma models admitting two
commuting KMA's classically. We have established their one loop
renormalizability and have given two independent arguments in
favour of the quantum integrability of the $x>1$ subset.

{\bf Acknowledegements}
We would like to thank M. Niedermaier and C. Zachos
 for calling
our attention to ref.\ [1] and ref.\ [14] respectively.
This work was partially supported by the
Hungarian National Science and Research Foundation
(Grant No.\ 2177 and No.\ 1815).
\vfill\eject
\centerline{\bf References}
\bigskip

\item{[1]} S. G. Rajeev, {\sl Phys. Lett.} {\bf B217} (1989) 123
\item{[2]} A. Polyakov and P. B. Wiegmann,
{\sl Phys. Lett.} {\bf B131} (1983) 121
\item{[3]} L. D. Faddeev and N. Yu. Reshetikin,
{\sl Ann. Phys.} {\bf 167} (1986) 227
\item{[4]} K. Pohlmeyer,
{\sl Comm. Math. Phys.} {\bf 46} (1976) 207
\item{[5]} L. D. Faddeev {\bf in} Les Houches lectures (1982)
on Recent Advances in Field Theory and Statistical Mechanics,
eds.\ J. B. Zuber and R. Stora (North Holland, Amsterdam, 1984)
\item{[6]} M. L\"uscher {\sl Nucl. Phys.} {\bf B135} (1978) 1
\item{[7]} H. De Vega, H. Eichenherr, J. M. Maillet,
{\sl Nucl. Phys.} {\bf B240} (1984) 377,
{\sl Comm. Math. Phys.} {\bf 92} (1984) 507
\item{[8]} D. Bernard, {\sl Comm. Math. Phys.} {\bf 137} (1991)
191
\item{[9]} V. E. Zakharov and A. V. Mikhailov,
{\sl Sov. Phys. JETP} {\bf 47} (1978) 1017
\item{[10]} A. Polyakov, {\sl Phys. Lett.} {\bf B72} (1977) 224
\item{[11]} A. B. Zamolodchikov  and A. B. Zamolodchikov,
{\sl Ann. Phys.} {\bf 120} (1979) 253
\item{[12]} D. Friedan, {\sl Ann. Phys.} {\bf 163} (1985) 318
\hfill\break
for further references see the recent review\hfill\break
V. V. Belokurov and D. I. Kazakov, {\sl Sov. J. Part. Nucl.}
{\bf 23} (1992) 577
\item{[13]} H. Osborn, {\sl Ann. Phys.} {\bf 200} (1990) 1
\item{[14]} Ch. Nappi, {\sl Phys. Rev.} {\bf D21} (1980) 418

\vfill
\end